# Quantum Resistance Standard Based on Epitaxial Graphene

Alexander Tzalenchuk[1], Samuel Lara-Avila[2], Alexei Kalaboukhov[2], Sara Paolillo[3], Mikael Syväjärvi[4], Rositza Yakimova[4], Olga Kazakova[1], T.J.B.M. Janssen[1], Vladimir Fal'ko[5], Sergey Kubatkin[2]

[1]*National Physical Laboratory, TW11 0LW Teddington, UK*

[2]*Department of Microtechnology and Nanoscience, Chalmers University of Technology, S-412 96 Göteborg, Sweden*

[3]*Department of Physics, Politecnico di Milano, 20133 Milano, Italy*

[4]*Department of Physics, Chemistry and Biology (IFM), Linköping University, S-581 83 Linköping, Sweden*

[5]*Physics Department, Lancaster University, Lancaster, LA1 4YB, UK*

**We report development of a quantum Hall resistance standard accurate to a few parts in a billion at 300 mK and based on large area epitaxial graphene. The remarkable precision constitutes an improvement of four orders of magnitude over the best results obtained in exfoliated graphene and is similar to the accuracy achieved in well-established semiconductor standards. Unlike the traditional resistance standards the novel graphene device is still accurately quantized at 4.2 K, vastly simplifying practical metrology. This breakthrough was made possible by exceptional graphene quality achieved with scalable silicon carbide technology on**

**a wafer scale and shows great promise for future large scale applications in electronics.**

The quantum Hall effect (QHE) is a fundamental phenomenon observed in two-dimensional electron systems with high charge mobility. It shows quantization[1] of transverse (Hall) resistance in rational fractions of the resistance quantum $R_K = h/e^2 = 25\,812.807\,572(95)\,\Omega$ – the von Klitzing constant[2]. The quantization of Hall resistance is accompanied by the vanishing of the dissipative longitudinal resistance, $R_{xx} = 0$. It is the result of the discreteness of the energy spectrum of two-dimensional electrons in a magnetic field – the Landau level quantization[3]. QHE offers a natural choice for the development of a resistance standard related to fundamental constants of nature: the electron charge $e$ and Planck's constant $h$. Because of its universality, the QHE resistance should be, in principle, independent of a particular embodiment of the two-dimensional material. Signatures of QHE were indeed observed in a number of materials, most recently in oxides[4]. Achieving resistance quantization, with the precision of few parts per billion required by metrology is, however, a much more complex and material-dependent issue. Suitability of material for QHE metrology depends on its homogeneity, stability, low intrinsic noise, possibility to attach low-resistance contacts to it[5]. Despite nearly 30 years of history, metrologically sound precision of the quantum Hall quantization at the level of parts per billion (ppb) was obtained only in two types of devices[6] – Si field effect transistors and GaAs heterostructures – two well-established materials of semiconductor microelectronics.

New materials are sought to expand the horizon of quantum metrology and to advance the understanding of QHE itself.

Graphene – a single layer of graphite – is a truly two-dimensional gapless semiconductor with electrons mimicking the behaviour of relativistic (Dirac) electrons[7]. This last feature of charge carriers in graphene is manifested most spectacularly through an unusual QHE[8]. In the quantum Hall regime the current is carried by a quantum state, spreading through the whole sample and the sequence of plateaux in the transverse resistance $R_{xy}$ depends on the topological (Berry) phase acquired by the charge moving in the magnetic field. This phase is zero in conventional materials, where $R_{xy} = \pm \frac{h}{ne^2}$ ($n \geq 1$); it is equal to $2\pi$ in bilayer graphene[9,10] leading to a sequence of QHE plateaux at $R_{xy} = \pm \frac{h/4e^2}{n+1}$ ($n \geq 0$), and $\pi$ in single-layer[11] with $R_{xy} = \pm \frac{h/4e^2}{n+1/2}$ ($n \geq 0$). The observation of this last sequence of QHE plateaux is therefore a smoking gun for the sample to contain monolayer graphene. Graphene should in principle be an ideal material for an implementation of a quantum resistance standard[12].

In reality, an impressive range of unconventional transport properties of electrons in graphene[13], including QHE, have been seen almost exclusively in flakes mechanically exfoliated from bulk graphite. Quantum Hall plateaux have been observed in graphene even at room temperature, albeit with an accuracy of 0.2%[14]. The highest experimentally achieved accuracy in exfoliated graphene flakes – 15 parts per million (ppm) at 300 mK[15] – is still modest by metrological standards. The main limiting factor is high contact resistance, $R_c \propto 1$ k$\Omega$, which induces additional noise leading to the involvement of higher Landau levels in the transport and limits the attainable accuracy. The small area of the flakes does not give much room for improvement in contact resistance. By the same token, the small channel width limits the measurement current while the small channel length prevents good equilibration of the edge states.

An alternative 'top-down' approach to produce graphene consists of growing it epitaxially[16]. Although angle-resolved photoemission studies of epitaxial graphene[17] have revealed an almost linear dispersion of carriers around the corners of a hexagonal Brillouin zone, $\varepsilon(p) = \pm vp$, and STM spectroscopy showed the sequence of Landau levels typical for graphene[18], Hall resistance quantization has not been observed in epitaxial graphene in contrast to exfoliated material despite numerous attempts. The difficulty was related to the lack of atomically accurate thickness control during film growth on the C-terminated facet, and most probably a strong variation of carrier density (doping) across the layers grown on the Si-terminated facet[19].

Here we report the first quantum Hall resistance standard based on large area epitaxial graphene synthesized on the Si-terminated face of silicon carbide. Graphene has become the third material ever to show Hall resistance quantization at a few ppb level at 300 mK and the first one to show robust quantization accurate to a few tens ppb at 4.2 K vastly simplifying practical metrology. This temperature is easily achievable with liquid helium or lately with pulse-tube coolers and is anyway required to cool most superconducting magnets.

The material studied in our experiments has been grown on the Si-terminated face of a 4H-SiC(0001) substrate[20]. The reaction kinetics on the Si-face is slower than on the

C-face because of the higher surface energy, which helps homogeneous and well controlled graphene formation[21]. Graphene was grown at 2000°C and 1 atm Ar gas pressure, which result in monolayers of graphene atomically uniform over more than 50 µm², as shown by low-energy electron microscopy. Twenty Hall bar devices of different sizes, from 160 µm x 35 µm down to 11.6 µm x 2 µm were produced on each 0.5 cm² wafer using standard electron beam lithography and oxygen plasma etching (Figure 1A-C). Atomic force microscopy (AFM) images reveal that the graphene layer covers the substrate steps like a carpet, preserving structural integrity. Contacts to graphene were produced by straightforward deposition of 3 nm of Ti and 100 nm of Au through a lithographically defined mask followed by lift-off, with the typical area of graphene-metal interface of $10^4$ µm² for each contact. This process favourably compares with a laborious contact preparation to two-dimensional electron gas in conventional semiconductor technology. Using low magnetic field measurements, we established that the manufactured material is *n*-doped, due to charge transfer from SiC[17,21], with the measured electron concentration in the range of $(5-8) \times 10^{11}$ cm$^{-2}$, mobility about 2400 cm²/Vs at room temperature and between 4000 and 7500 cm²/Vs at 4.2 K, almost independent of device dimensions and orientation with respect to the substrate terraces. Here we concentrate on the properties of the three smallest and the two largest devices.

Figure 2A shows typical behaviour for the manufactured devices of the longitudinal (dissipative) $R_{xx}$ and the transverse (Hall) $R_{xy}$ resistance of a 2 µm wide Hall bar at 4.2 K and magnetic field up to 14 T. At high magnetic field we clearly identify two QHE plateaux, at $R_{xy}^{(0)} = R_K/2$ (n=0) and $R_{xy}^{(1)} = R_K/6$ (n=1) corresponding to the filling factors $\nu = 2$ and $\nu = 6$ respectively. In graphene $\nu = 2$ corresponds to the fully occupied zero-energy Landau level (n=0) characterized by the largest separation $v\sqrt{2\hbar eB/c}$ from other Landau levels in the spectrum and hence the Hall resistance quantization is particularly robust. This plateau appears in the field range of 9-12 T, depending on the carrier concentration (which is beyond our control in this experiment) and is accompanied by a vanishing $R_{xx}$. Stronger doping leads to a higher onset field. The $n=1$ plateau, at $v=6$, is not so flat, and $R_{xx}$ develops only a weak minimum. There is also a trace of a structure corresponding to ν = 10. Their presence confirms that the studied material is indeed monolayer graphene. At low magnetic fields we observe Shubnikov-de Haas oscillations as well as characteristic features of phase coherence of electrons in a disordered conductor: a weak localization peak and reproducible conductance fluctuations within the interval $|B|< 2$ T.

Low contact resistance, complete vanishing of the longitudinal resistance on the plateau and a reasonably high current, that the contact and the material itself can sustain before QHE breaks down, are critical ingredients of a quantum Hall resistance standard. Large area epitaxial graphene makes it possible to achieve all these objectives. We chose the largest available sample, 160 μm long and 35 μm wide, for precision measurements. In this sample, the resistance of contacts $R_c$ to the graphene layer determined at the plateau (Figure 2B) was about 1.5 Ω (cf. $R_c \propto 1$ kΩ for exfoliated graphene). $R_{xx}$ stayed zero for bias currents up to 5 μA at 4.2 K, increasing to 13 μA at 300 mK (Figure 2C).

The accuracy of Hall resistance quantization in graphene was established in measurements traceable to the GaAs quantum Hall resistance standard. Direct comparison of the GaAs and the graphene samples is difficult to arrange as they both require cryogenic temperatures, but different magnetic fields. We therefore used a thoroughly characterized 100 Ω resistor thermally stabilized at room temperature as a transfer standard. The measurements on the graphene and the semiconductor samples were done only one week apart such that drift in the transfer standard was negligible (<< ppb). The comparison was performed using a cryogenic current comparator (CCC) bridge with the overall relative measurement uncertainty below 1 ppb[22]. Figure 3A shows how the mean relative deviation of $R_{xy}^{(0)}$ from $R_K/2$ depends on the measurement current through the graphene device. The optimal conditions at 300 mK were obtained for a bias of 11.5 µA, 15% below the breakdown current established in the measurements of $R_{xx}$. The quantization accuracy (mean relative deviation of 129 measurements ± standard error of the mean) +0.4 ± 3 parts in $10^9$, inferred from our measurements is a 4 order of magnitude improvement on the previous best estimate achieved in an exfoliated graphene sample. The reported result readily puts epitaxial graphene quantum Hall devices in the same league as their semiconductor counterparts. Note that our result was obtained on a sample which is substantially smaller than the semiconductor devices used for calibration and without any optimization. Furthermore, from Figure 3A it can be seen that graphene is still accurately quantized at 4.2 K. At this temperature the measurement current has to be reduced to 2.3 µA and consequently gives a larger uncertainty over a comparable time interval. Nevertheless, these measurements demonstrate that $R_{xy}^{(0)}$ in epitaxial graphene is accurately quantized and can be used for precision resistance metrology even at elevated temperatures.

In order to demonstrate convergence of the measurement process and to see what kind of noise limits the precision of our measurements, in Figure 3B we plot the Allan deviation[23,24] of $R_{xy}^{(0)}$ from $R_K/2$ against the measurement time τ. The data plotted in such way can be accurately fitted by $1/\tau^{1/2}$ dependence – the behaviour typical of the predominantly white, i.e., random and uncorrelated, noise. This justifies the use of the standard measures of uncertainty and suggests that even these very accurate results can be further improved if one is prepared to measure for longer.

As seen in the AFM images the graphene Hall bars are patterned across many substrate terraces. Our measurements of QHE reveal that the coherent transport and the chirality of charge carriers inherent to pristine graphene are preserved notwithstanding. These observations are intriguing and deserve more attention – apparently interactions with the substrate lattice and its defects, such as terrace edges (Figure 1A&B), make no negative effect on the graphene properties. Further studies are needed to clarify these issues. In any case, structural integrity and uniformity of epitaxial graphene film over hundreds of micrometers, decent reproducibility of parameters, such as mobility and concentration of charge carriers, across a half-centimetre wafer and wafer-to-wafer achieved in this work already speak volumes in favour of the prospects of the SiC technology for microelectronics applications possibly reaching far beyond quantum metrology.

To summarize, we report quantization of Hall resistance peculiar to graphene in devices produced on the Si-terminated surface of SiC. Having observed the QHE in several devices produced on distant parts of a single large-area wafer and established the Hall resistance quantization with a standard deviation of only few parts per billion at 300 mK, we can confirm that material synthesized on the Si-terminated face of SiC offers a suitable platform for implementation of the quantum resistance standard.

Reduced noise and higher breakdown current at even lower temperatures may enable further improvement of metrological precision. In the future, this may advance the understanding of the QHE effect itself, by finding whether there exist measurable deviations of the quantized Hall resistance from the universally accepted value of $R_K$. But even more importantly, here, graphene passed the definitive QHE test in devices over hundred micrometers long manufactured by patterning a centimetre-size chip, which opens up bright prospects for scalable electronics based on graphene.

**Acknowledgement:** Authors wish to thank Lars Walldén, Tomas Löfwander, John Gallop and Tord Claeson for stimulating discussions, Stepehen Giblin and Jonathan Williams for help with experiments. We are grateful to NPL Strategic research programme, Swedish SSF and VR for financial support.

**Correspondence** and requests for materials should be addressed to AT:

alexander.tzalenchuk@npl.co.uk

Figure 1. Sample layout and structure.

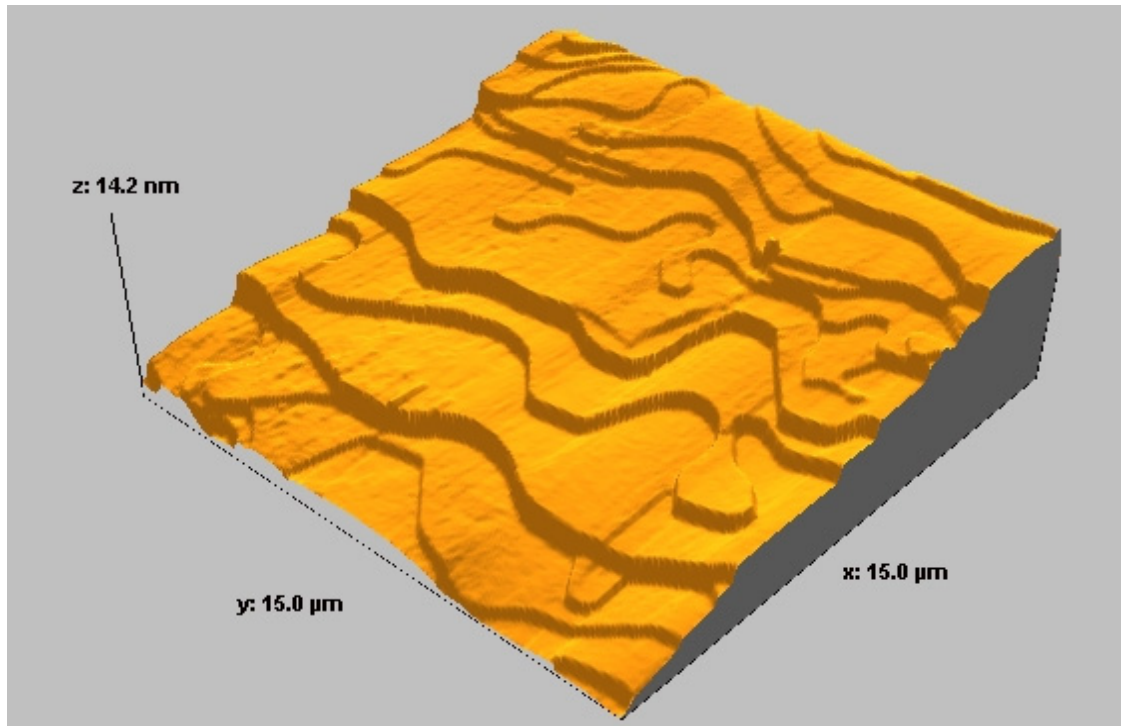

A. Atomic force microscopy (AFM) images of the sample: large flat terraces on the surface of Si-face of 4H-SiC(0001) substrate with graphene after high-temperature annealing in Ar atmosphere.

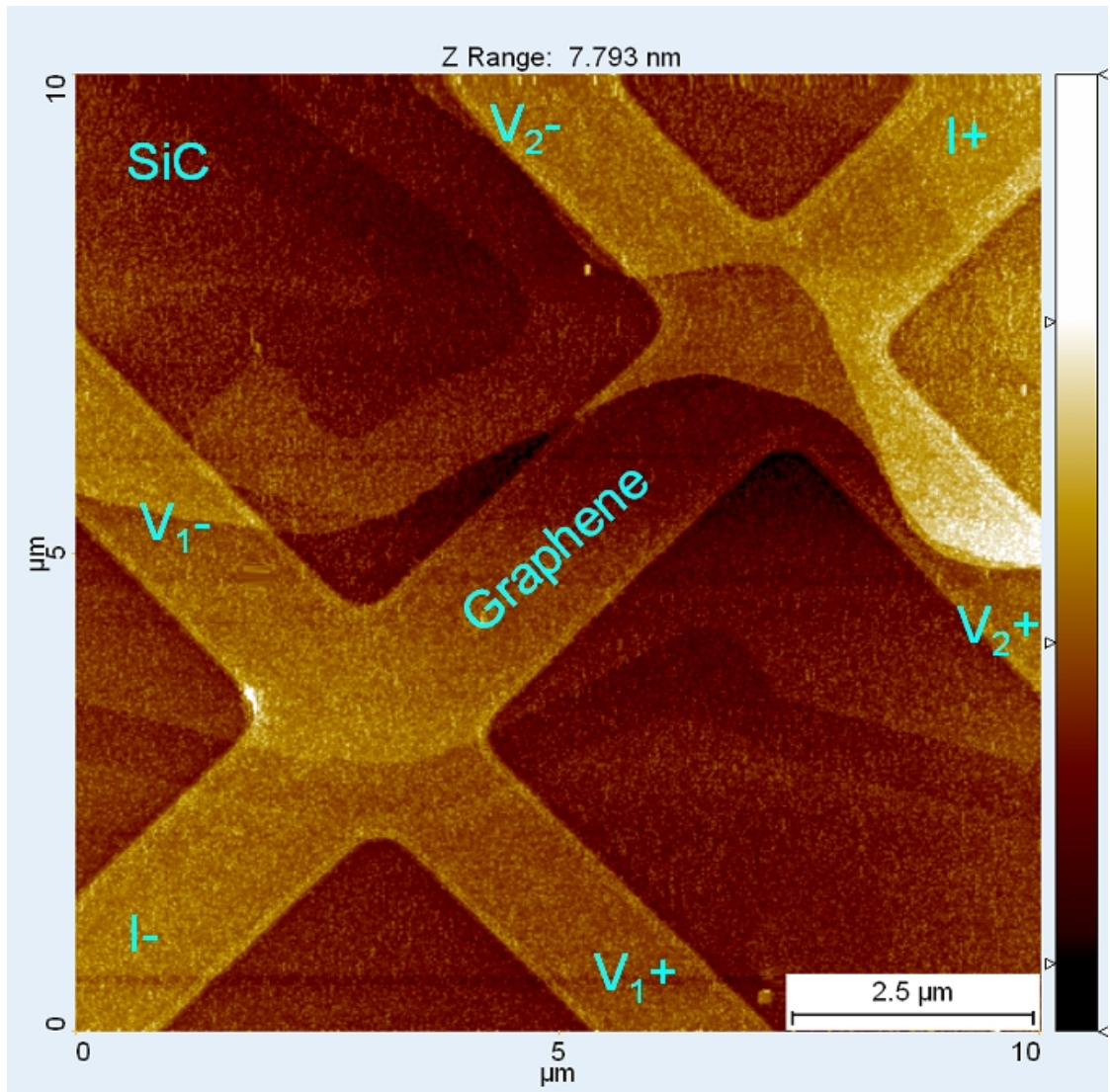

B. Graphene patterned in the nominally 2 μm wide Hall bar configuration on top of the terraced substrate.

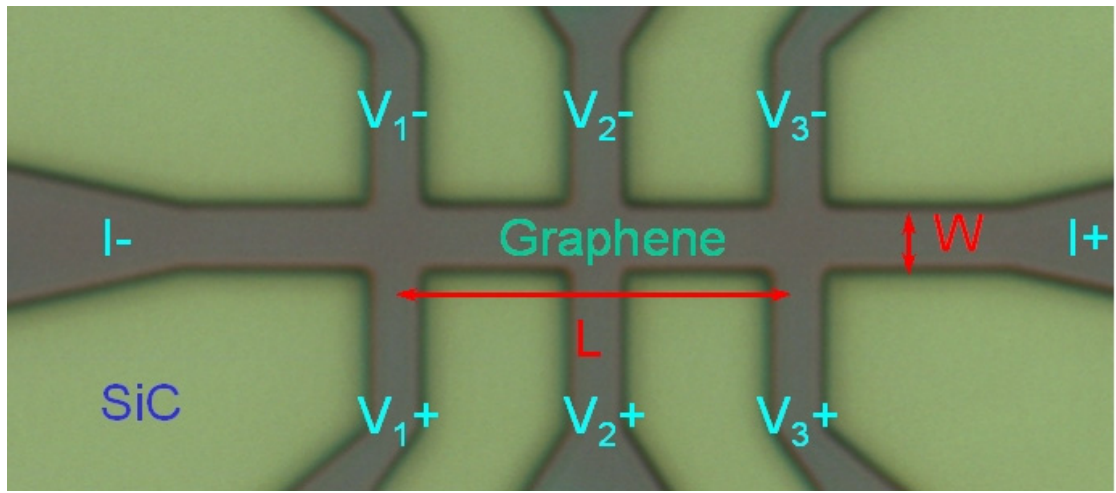

C. A device with length L=11.6 μm and width W=2 μm. To visualize the Hall bar this optical micrograph was taken after oxygen plasma treatment, which formed the graphene pattern, but before the removal of resist.

Figure 2. Quantum Hall effect in epitaxial graphene.

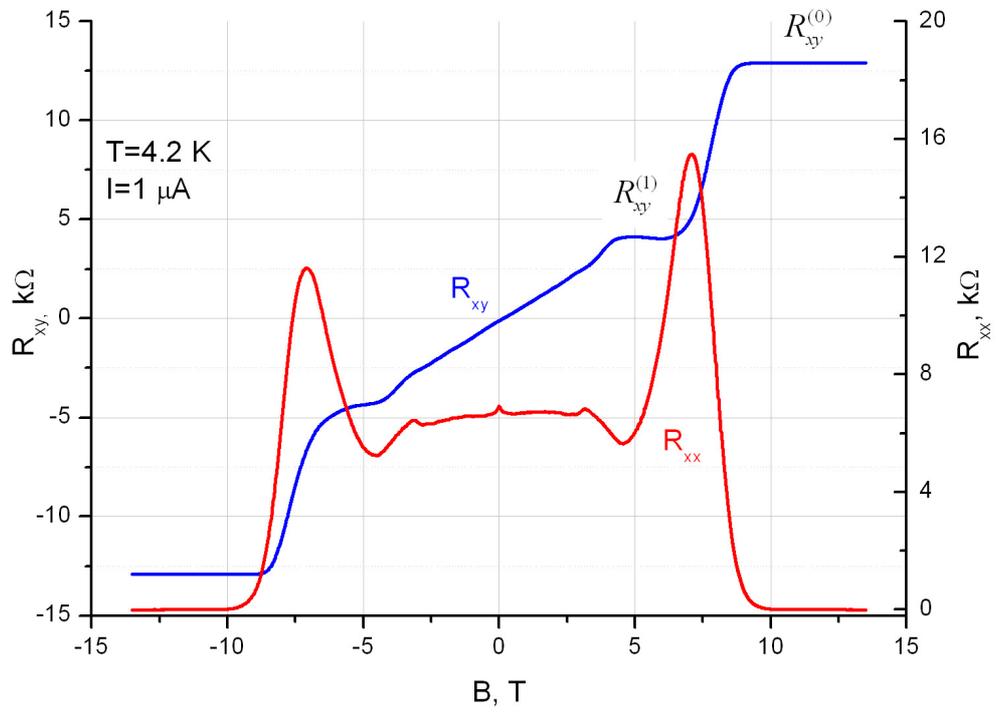

A. Transverse ($R_{xy}$) and longitudinal ($R_{xx}$) resistance measured on the 2 μm wide device at T=4.2 K across the pairs of contacts ($V_1$+$V_1$-) and ($V_2$+ $V_1$+) respectively with 1 μA current between I+ and I-.

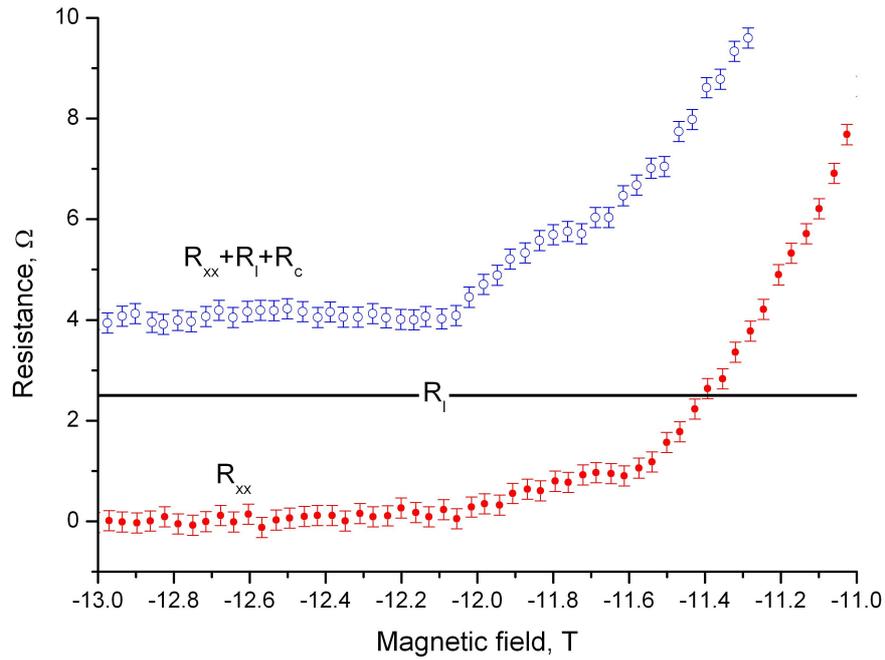

B. Measurements of the longitudinal resistance $R_{xx}$ performed in four-point configuration (red), which excludes contact resistances, and in three-point configuration, which apart from $R_{xx}$ include the contact resistance $R_c$ and the resistance of the leads from room temperature electronics down to the sample $R_l$. On the plateau $R_{xx}$ vanishes, $R_l = 2.5\ \Omega$ and $R_c$ appears to be about 1.5 $\Omega$ for all measured contacts – of the same order as good contacts to GaAs samples and ~1000 times lower than contacts to exfoliated graphene.

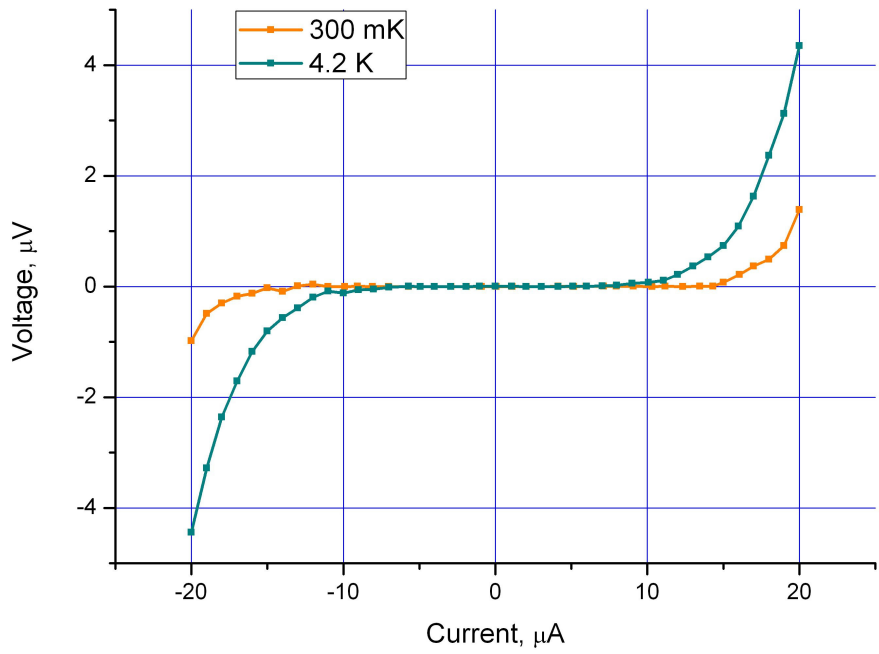

C. Determination of the breakdown current $I_{max}$ of non-dissipative transport from measurement of the current-voltage characteristic in the longitudinal direction at 14 T. $I_{max} \approx 13$ µA at 300 mK and $I_{max} \approx 5$ µA at 4.2 K.

Figure 3. Determination of the Hall resistance quantization accuracy.

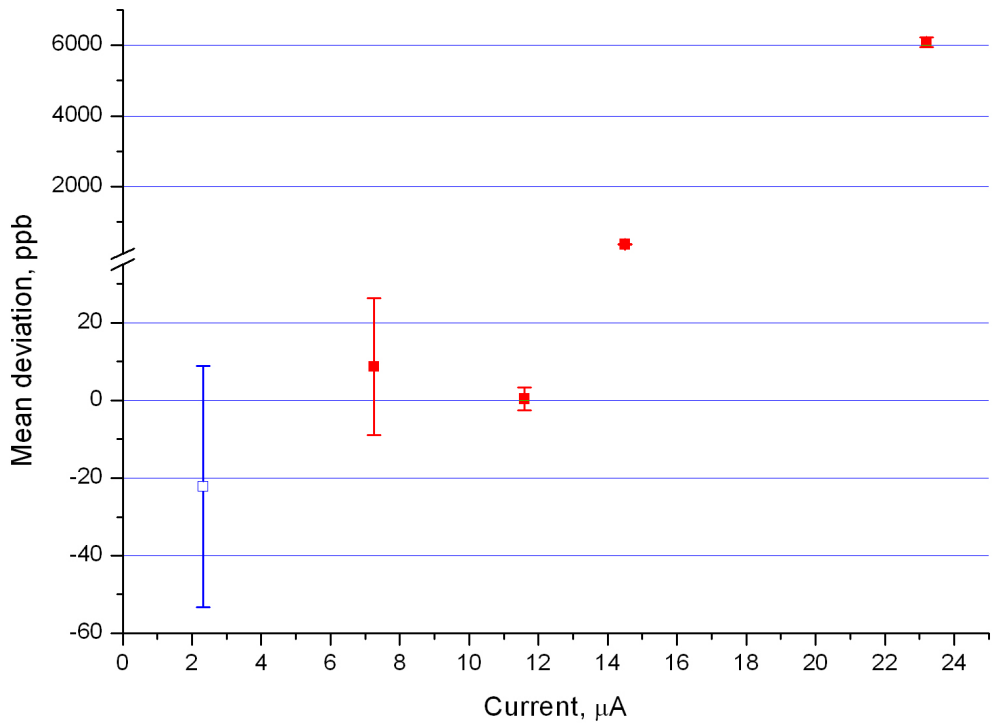

3A. Mean relative deviation of $R_{xy}^{(0)}$ from $R_K/2$ at different bias currents (ppb=parts per billion). The value at the smallest current was measured at 4.2 K (blue open square), all other values – at 300 mK (red solid squares).

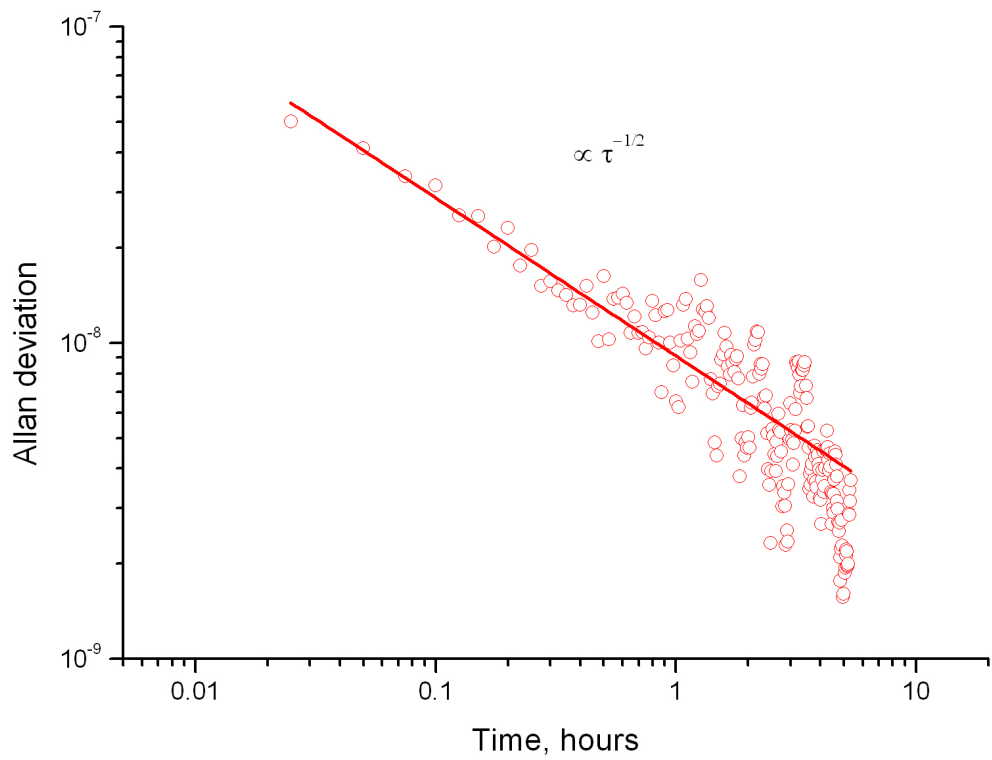

3B. Allan deviation of $R_{xy}^{(0)}$ from $R_K/2$ vs. measurement time $\tau$. The square root dependence indicates purely white noise.